%% file: iclr2024_conference.tex
\documentclass{article} 
\usepackage{iclr2024_conference,times}

\input{math_commands.tex}

\usepackage{hyperref}
\usepackage{url}
\usepackage{multirow}
\usepackage[T1]{fontenc}
\usepackage{newtxmath, newtxtext}

\title{Building a Luganda Text-to-Speech Model From Crowdsourced Data}

\author{Sulaiman Kagumire \& Andrew Katumba \\
Department of Electrical and Computer Engineering \\
Makerere University\\
Kampala, Uganda \\
\texttt{\ sulaiman.kagumire@gmail.com, andrew.katumba@mak.ac.ug} \\
\And
Joyce Nakatumba-Nabende \\
Department of Computer Science \\
Makerere University\\
Kampala, Uganda \\
\texttt{\ joyce.nabende@mak.ac.ug} \\
\AND
John Quinn \\
Sunbird AI \\
Kampala, Uganda \\
\texttt{jq@sunbird.ai}
}

\iclrfinalcopy 
\begin{document}

\maketitle

\begin{abstract}
Text-to-speech (TTS) development for African languages such as Luganda is still limited, primarily due to the scarcity of high-quality, single-speaker recordings essential for training TTS models. Prior work has focused on utilizing the Crowdsourced Common Voice Luganda recordings of multiple female speakers aged between 20-49. Although the generated speech is intelligible, it is still of lower quality compared to their model trained on studio-grade recordings. This is due to the insufficient data preprocessing methods applied to improve the quality of the Common Voice recordings. Furthermore, speech convergence is more difficult to achieve due to varying intonations from multiple speakers, as well as background noise in the training samples. In this paper, we show that the quality of Luganda TTS from Common Voice can improve by training on multiple speakers of close intonation in addition to further preprocessing of the training data. Specifically, we selected six female speakers with close intonation determined by subjectively listening and comparing their voice recordings. In addition to trimming out silent portions from the beginning and end of the recordings, we applied a pre-trained speech enhancement model to reduce background noise and enhance audio quality. We also utilized a pre-trained, non-intrusive, self-supervised Mean Opinion Score (MOS) estimation model to filter recordings with an estimated MOS over 3.5, indicating high perceived quality. Subjective MOS evaluations from nine native Luganda speakers demonstrate that our TTS model achieves a significantly better MOS of 3.55 compared to the reported 2.5 MOS of the existing model. Moreover, for a fair comparison, our model trained on six speakers outperforms models trained on a single-speaker (3.13 MOS) or two speakers (3.22 MOS). This showcases the effectiveness of compensating for the lack of data from one speaker with data from multiple speakers of close intonation to improve TTS quality.
\newline
Text-to-Speech, Crowdsourced, Luganda, Speech, Text, Multiple Speakers 

\end{abstract}

\section{Introduction}
Speech synthesis technology has significantly advanced in recent years with deep neural network models, allowing for successful applications such as speech-based virtual assistants. While these advancements are evident in popular languages such as English, French, and Spanish, many of the low-resourced African languages are still lagging far behind. Notably, major Text-to-Speech (TTS) platforms like Amazon's Polly, NaturalReader, Voice Dream Reader, Speechify, and Google's Speech Recognition and Synthesis, have yet to include support for various African languages. The disparity originates from the fact that TTS models deployed in such platforms require relatively large numbers of high-quality recordings with text transcriptions from a single professional speaker to generate natural-sounding audio \cite{ogayo}. For many low-resourced African languages such as Luganda, not only is this type of data difficult to collect but also the existing ones are often proprietary, which limits its accessibility for TTS development. Luganda is a Bantu language spoken by more than 20 million people globally and by over 16.7\% of Uganda's population \cite{babirye}.  Despite having a relatively large number of speakers, the language is arguably overlooked in the development of current TTS systems due to the the lack of high-quality single-speaker TTS corpora.  

To reduce the dependency on single speaker datasets in low-resource settings, previous studies \cite{effects, luong} have shown that training TTS models on a mixture of available multiple speakers can generate synthetic speech with better quality and stability than a speaker-dependent one. A recent work \cite{sunbird} has developed a Luganda TTS model\footnote{\url{https://huggingface.co/Sunbird/sunbird-lug-tts-commonvoice-female}} from crowdsourced Common Voice Luganda recordings of multiple female speakers aged 20-49. Despite the acceptable intelligibility of their model, it is of lower quality compared to their model trained on single-speaker studio recordings\footnote{\url{https://huggingface.co/Sunbird/sunbird-lug-tts}}. Additionally, speech convergence was more difficult to attain with Common Voice due to the varying intonations from multiple speakers, as well as the existence of background noise in the training data because Common Voice is primarily collected for Automatic Speech Recognition (ASR), rather than speech generation \cite{common}. There is therefore a need to enhance TTS models trained on the crowdsourced Common Voice Luganda dataset.

In this paper, we present an improved Luganda TTS model trained on the Common Voice recording of six female speakers with close intonation that is determined through listening and comparing the voices. This approach upgrades speech convergence during synthesis, offering a noticeable improvement over the existing model where individual training voices are quite diverse. On top of trimming silent parts from the beginning and end of each recording, a technique adopted by the existing model, we describe additional data preprocessing steps to enhance the quality of the training speech and text data. These measures not only refine the quality of the training speech data but also improve the generated speech quality. By fine-tuning an English pre trained and end-to-end TTS model on the crowdsourced Common Voice recordings, our model achieves a level of speech naturalness that surpasses that of the existing model. Furthermore, we also compare the performance of the model trained on six female speakers to models trained on a single speaker or two speakers. Evaluation results show that training on a more extensive and diverse dataset sounds more natural than those trained on single speaker or fewer speakers. 

By incorporating speech data from multiple speakers with close intonation, our model became more robust to common speech synthesis challenges such as varied pronunciation styles and incorrectly labeled sentences. The model captured a broader range of speech nuances, such as subtle variations in pitch, tone, and emotion, which are characteristics of natural speech. Additionally, by exposing the model to a range of phonetic and prosodic patterns within a controlled intonation range, its ability to generalize and accurately reproduce speech sounds for unseen text inputs was significantly improved.

In Section 2 we discuss related work that utilizes crowdsourced data for Luganda TTS. Section 3 describes the methodology, including the data and the preprocessing techniques. Section 4 presents the Experiments and subjective evaluation results of our model. Section 5 concludes the paper. 

\newpage
\section{Related Work}
\label{related}
There is previous effort towards building TTS models using ASR crowdsourced data as an alternative to studio-quality datasets. By leveraging the Common Voice English dataset, a multi-speaker GlowTTS model \cite{canwe} was trained on recordings automatically selected  using a non-intrusive mean opinion score (MOS) estimator, WVMOS\footnote{\url{https://github.com/AndreevP/wvmos}}. Their approach improves the overall quality of generated utterances by 1.26 MOS point with respect to training on all the samples and by 0.35 MOS point with respect to training on the LibriTTS dataset\footnote{\url{https://openslr.org/60/}}.

In their recent work \cite{sunbird}, Sunbird AI\footnote{\url{https://sunbird.ai/}} presented a Tacotron2-based model fine-tuned on 15,000 Common Voice Luganda recordings from female speakers aged between 20 and 49. They employed voice activity detection\footnote{\url{https://github.com/wiseman/py-webrtcvad/blob/master/example.py}} to trim off the silent portions from the beginning and end of each speech recording. Whilst their model's generated speech is intelligible enough to be usable in some practical applications, it is of lower quality due to existence of background noise in the training data. The model learnt that the noise is an intrinsic part of the speech signals it needs to generate. As a result, the model's output includes similar noise, lowering the overall quality of the generated speech.

\section{Methodology}
\label{metholody}
\subsection{Data}
In this work, we utilized the free crowdsourced monolingual Luganda (version 12.0) speech data obtained from the Mozilla Common Voice platform\footnote{\url{https://commonvoice.mozilla.org/en}}. The data contains mp3 audio recordings which are up-voted or down-voted by volunteers according to a list of criteria\footnote{\url{https://commonvoice.mozilla.org/en/guidelines}}. Utterances with more than two up-votes are marked as validated, otherwise, they are marked as invalidated. The data comes with a CSV file which contains important information about the audio files including the text transcription of each speaker's utterance, name/path of the utterance, gender and age of the speaker. Throughout all experiments, we only considered validated utterances from female speakers.
\subsection{Speaker Selection}
Among the top 20 contributing female speakers, we identified six voices of speakers with closely matching intonation (the rise and fall of a voice in speaking). We determined closely similar speakers by listening to and comparing the intonation of their respective recordings. This similarity was essential for ensuring that the synthesized speech sounds natural and cohesive, as it minimizes the variability that arises when combining speech data from multiple speakers.

To assess the impact of utilizing multiple speakers over a single speaker or a smaller group of speakers, we trained the model using data from the highest contributing female and then separately with data from the two highest contributing females out of the six selected speakers. Table~\ref{tab:1} provides the statistical details for the six speakers, alongside the single speaker and the two speakers.
\begin{table}[h]
\centering
\caption{ Statistics of the six speakers, one speaker and two speakers}
\label{tab:1}
\begin{tabular}{ccccccl}
\hline
\multirow{2}{*}{ \bf LANGUAGE} & \multirow{2}{*}{\bf SPEAKERS} & \multicolumn{3}{c}{\bf STATISTICS} \\
\cline{3-5}
& & \bf MAX LENGTH & \bf MIN LENGTH& \bf DURATION (hrs) \\
\hline
\multirow{3}{*}{Luganda} & one & 8.49&1.23 &8.06  \\
& two  &8.49 &1.17 &10.17   \\
& six &10.72 &1.14 &19.04   \\
\hline
\end{tabular}
\end{table}

\subsection{Data Preprocessing}
\subsubsection{Audio Quality Enhancement}
A major problem with Common Voice speech data is the existence of distorted or noisy audio samples. The presence of mouse clicks, low-frequency noise, background speakers and music, among others, can be observed. In addition, silences at the beginning and end of several utterances can eb noticed, which causes misalignment between text and audio that degenerates TTS quality.

To eliminate silence at the start and end of the recordings,  we utilized WebRTC Voice Activity Detection (VAD)\footnote{\url{https://github.com/wiseman/py-webrtcvad/blob/master/example.py}} that was used in the previous work by Sunbird AI. First, each wav file is read, and its audio content is processed in frames of 30 milliseconds duration using a VAD algorithm provided by the webrtcvad\footnote{\url{https://github.com/wiseman/py-webrtcvad}} library. The VAD algorithm identifies segments of the audio containing speech. Once speech segments are detected, they are collected and concatenated to form continuous speech segments. These speech segments are then saved as new WAV files in a designated directory. Additionally, any speech segments shorter than 1 second are considered problematic and marked for further examination. This process effectively filters out silent parts of the audio files, retaining only segments containing speech.

Furthermore, to enhance the quality of the trimmed audios, we applied a pre-trained speech enhancement model\footnote{\url{https://github.com/facebookresearch/denoiser}} that works directly on the raw audio data to denoise and enhance the audio quality. The model is based on an encoder-decoder architecture with skip-connections \cite{defos}. It is optimized on both time and frequency domains, using multiple loss functions such as the regression loss function (L1 loss), complemented with a spectrogram domain loss \cite{parallel, probability}. Empirical evidence shows that it is capable of removing various kinds of background noise including stationary and non-stationary noises, as well as room reverb \cite{defos}. 

Additionally, to ensure inclusion of only good-quality training audio samples, we considered an absolute objective speech quality measure based on direct MOS score prediction by a fine-tuned wave2vec2.0 model (WV-MOS) \cite{andreev}. WV-MOS is reported to have better system-level correlation with subjective quality measures than the other objective metrics such as Perceptual Evaluation of Speech Quality (PESQ) \cite{pesq} and Short-Time Objective Intelligibility (STOI) \cite{stoi}. We selected denoised recordings with an estimated MOS score above 3.5 by WV-MOS\footnote{\url{https://github.com/AndreevP/wvmos}}.
\subsubsection{Text Preprocessing}
We eliminated all transcripts with less than three words, alongside their corresponding audio samples. We also replaced special characters with standard representations compatible with TTS input. For example, the \textbf{ŋ} character in Luganda was replaced with \textbf{ng} characters as shown in Table ~\ref{tab:2}. Additionally, incorrectly written punctuations were also replaced with standard punctuation symbols. For example: double or three full stops were found in the middle or end of the sentence, and these were replaced with one full stop.

\begin{table}[h]
\centering
\caption{Sample of a preprocessed sentence with the ŋ charater in the original sentence replaced by the ng characters in the new sentence }
\label{tab:2}
\begin{tabular}{cc}
\hline
\textbf{ORIGINAL SENTENCE} & \textbf{NEW SENTENCE} \\
\hline
yantuma \textbf{ŋende} ndeete ekidomola ky'amazzi. & yantuma \textbf{ngende} ndeete ekidomola ky'amazzi. \\
\hline
\end{tabular}
\end{table}
\subsection{Model}
For this work, we fine-tuned a Variational Autoencoder with Adversarial Learning (VITS) TTS model \cite{vits} pre-trained on the LJ Speech, a single-speaker English dataset \cite{ljspeech17}. VITS is an end-to-end speech synthesis model that predicts a speech waveform conditional on an input text sequence. It is a conditional variational autoencoder (VAE) \cite{vae} comprised of a posterior encoder, decoder, and conditional prior encoder.

VITS is trained end-to-end with a combination of losses derived from variational lower bound and adversarial training. To improve the expressiveness of the model, normalizing flows are applied to the conditional prior distribution. During inference, the text encodings are up-sampled based on the duration prediction module and then mapped into the waveform using a cascade of the flow module and HiFi-GAN decoder. VITS's end-to-end approach combines the GlowTTS encoder \cite{glow} and HiFiGAN vocoder \cite{hifi}, which simplified our TTS training pipeline, making it more streamlined and easier to implement.
\subsection{Evaluation}
To assess the performance of our model trained on six speakers, we conducted mean opinion score (MOS) tests. These tests involved synthesizing audio samples using each model and subsequently evaluating their perceived naturalness on a 5-point scale. We generated 100 audio samples from the model and presented them to 10 native Luganda speakers, who were tasked with evaluating the naturalness of the synthesized speech.

The MOS is calculated as:
\begin{equation*}
    \text{MOS} = \frac{\sum \text{scores}}{N}
\end{equation*}

Where:
\begin{itemize}
  \item MOS: Mean Opinion Score.
  \item $\sum$ scores: Sum of individual scores provided by all raters.
  \item $N$: Total number of individual scores provided by all raters.
\end{itemize}

\section{Experiments and Results}
Model training experiments were carried out using the Coqui TTS Library\footnote{\url{https://github.com/coqui-ai/TTS}}. For all experiments, we fine-tuned an English VITS model provided by Coqui. All models were trained on a single NVIDIA GeForce GTX 1080 Ti GPU for 1,000,000 steps at a learning rate of 0.001 and a batch size of 16. The AdamW optimizer was used to update the model’s parameters with a weighting decay of 0.01. The dataset was split with a 90\% training set and a 10\% validation set. The best training parameter settings that were found to give the best possible synthesis performance on the Common Voice data are shown in Table~\ref{tab:3}. These underwent a couple of quality checks including examining for the noise level of the clips by checking spectrograms as well as finding suitable audio processing parameters. We also reduced the sample rate to 22.05 kHz to speed up data-loading and consequently accelerate model training.
\begin{table}[h]
\centering
\caption{Hyperparameters tuned to adapt the pre-trained VITS model to Common Voice} 
\begin{tabular}{cc}\hline
\bf HYPERPARAMETER & \bf VALUE\\\hline
preemphasis & 0.98 \\
  ref\_level\_db& 20\\
  mel\_fmax & 8000 \\
    log\_func& np.log \\
 spec\_gain & 1 \\
 use\_phoneme&False \\
 phoneme\_language&False \\ \hline
\end{tabular}
\label{tab:3}
\end{table}

MOS evaluation results by ten native Luganda speakers of our model are shown in Table~\ref{tab:4}. We compare our results to the existing model MOS results as reported in their paper \cite{sunbird}. They fine-tuned a Tacotron2 model on Luganda Common Voice of female speakers within the age range of 20-49. Despite the existing model being trained on a broader range of speakers, our model achieved a higher quality score.
\begin{table}[h]
\centering
\caption{MOS comparison of our VITS model trained on six female Luganda speakers to the existing Tacotron2 model} 
\label{tab:4}
\begin{tabular}{cc}\hline
\textbf{MODEL} & \textbf{MOS}\\\hline
 Tacotron2-based (existing) & 2.50 \\
VITS-based (ours)&\textbf{3.55} \\ \hline
\end{tabular}
\end{table}
Our results highlights the importance of not just the quantity but also the selection of speakers with close intonation. Our approach focused on selecting speakers with closely aligned intonations, which proved to be more effective for maintaining a more consistent synthesis voice than the broader age-based selection used in the Tacotron2 model. Additionally, the quality of our model is influenced by the data preprocessing steps taken to clean the training speech recordings. By reducing background noise and selecting recording with an estimated MOS of above 3.5, we ensure that the model learns from higher quality speech samples than only trimming silents parts, which allows distorted samples in the training data. This approach allows our model to concentrate more effectively on learning the characteristics of the speech itself, resulting in more natural-sounding synthesized speech.
\subsection{Comparison by Number of Speakers}
We also trained two models on a single speaker and two speakers with the highest number of samples from the six speakers. As shown in Table~\ref{tab:5}, we compare their performance to our model trained on all the six speakers, which performs better. We also noticed that the model trained on two speakers of close intonation as well performed better than the one trained on a single speaker.
\begin{table}[h]
\centering
\caption{MOS comparison by different number of speakers}
\begin{tabular}{cc}
\hline
\textbf{SPEAKERS} & \textbf{MOS}  \\
\hline
One & 3.13  \\
Two & 3.22   \\
Six & \textbf{3.55} \\
\hline
\end{tabular}
\label{tab:5}
\end{table}
Our results highlight the significant impact of utilizing multiple speakers over a single speaker or fewer speakers. The observed improvement in naturalness with multiple speakers can be attributed to the benefits of training on a more extensive and diverse dataset. The incremental improvements in MOS from one to two speakers and then to six speakers suggest that while even a small increase in speaker diversity can enhance speech quality, larger and more diverse datasets offer more significant improvements. By incorporating speech data from many speakers with close intonation, the model became more robust against common speech synthesis challenges such as varied pronunciation styles and incorrectly labeled sentences. This means that training on multiple speakers of close intonation compensates for the lack of data from a single speaker, a common challenge for underrepresented languages in speech synthesis research. The broader range of speech samples from speakers provides the model with a comprehensive understanding of the language's phonetic and prosodic nuances, enabling it to generalize better to new inputs and maintain a more consistent voice quality during synthesis.
\section{Conclusion and Future Work}
In this paper, we presented a Luganda TTS model trained from crowdsourced Common Voice data of multiple speakers of closely matching intonation determined through listening and comparing their voices. Additionally, we presented preprocessing techniques aimed at reducing noise from Luganda Common Voice speech and text data, crucial for effective TTS training. MOS evaluations results show that our model (3.55 MOS) trained on six speakers of similar intonation can produce better quality than the existing model (2.50 MOS) that was trained on multiple speakers based on age range. Our results show that by carefully selecting multiple speakers of close intonation, the model can generate a more consistent and better quality voice at synthesis than training on speakers of different intonations despite being from a certain age range.  Furthermore, our results highlight the importance of training on a diverse and multi-speaker dataset, as it consistently produces better quality outputs compared to models trained on a single speaker or fewer speakers. We show that this is true for our model trained on six speakers and for models trained on a single speaker and two speakers. Future work will focus on applying our methodology to build models from Common Voice data of other low-resourced African languages.


\bibliography{iclr2024_conference}
\bibliographystyle{iclr2024_conference}


\end{document}

%% file: math_commands.tex
\usepackage{amsmath,amsfonts,bm}

\def\eqref#1{equation~\ref{#1}}

\def\1{\bm{1}}

\DeclareMathAlphabet{\mathsfit}{\encodingdefault}{\sfdefault}{m}{sl}
\SetMathAlphabet{\mathsfit}{bold}{\encodingdefault}{\sfdefault}{bx}{n}

